\documentclass[review]{elsarticle}
\usepackage{hyperref}

\journal{IEEE}

\begin{document}

\begin{frontmatter}
\title{Towards Better Problem Finding and Creativity in Graduate School Education\tnoteref{t1}}
\tnotetext[t1]{This work has been submitted to the IEEE for possible publication. Copyright may be transferred without notice, after which this version may no longer be accessible.}
\author[hu-eng]{Ankit A. Ravankar\corref{cor2}}
\ead{ankit@eng.hokudai.ac.jp}
\author[hu-ed]{Shotaro Imai}
\author[kt-eng]{Abhijeet Ravankar\corref{cor1}}
\author[hu-agr]{Tomomichi Kato\corref{cor2}}
\author[hu-agr]{Taichi Takasuka\corref{cor2}}
\author[hu-ed]{Teruyuki Tsuji}
\author[hu-ed]{Kaori K. Shigetomi}
\author[hu-ed]{Ken Saito}
\cortext[cor2]{Principal corresponding author}
\address[hu-eng]{Research Faculty of Engineering, Hokkaido University, Sapporo, Japan 060-8628}
\address[hu-ed]{Institute for the Advancement of Higher Education, Hokkaido University, Sapporo, Japan}
\address[hu-agr]{Research Faculty of Agriculture, Hokkaido University, Sapporo, Japan}
\address[kt-eng]{Research Faculty of Engineering, Kitami Institute of Technology, Kitami, Japan}

%
%
%

\begin{abstract}
The current graduate school education system has largely been focusing on producing better learners and problem solvers. The rise of problem based learning approaches are testimonial to the importance of such skills at all levels of education from early childhood  to graduate school level. However, most of the programs so far have focused primarily on producing better problem solvers neglecting problem finding at large. Problem finding, an important skill is a subset and first step in creative problem solving. Most studies on problem finding skills have only focused on industries and corporations for training employees to think out of the box for innovative product design and development. At school or university level, students are generally given a well-defined problem in most Problem Based Learning (PBL) scenarios and problem discovery or how to deal with ill-structured problems is mostly ignored.  In this study, we present the Nitobe School Program and discuss our unique curriculum to teach problem finding in graduate school education. We show how introducing problem finding at graduate level increases student's ability to comprehend difficult and wicked problems in a team based learning environment. Moreover, we present how it influences creativity in graduate students resulting in better problem solvers.  
\end{abstract}


\end{frontmatter}


\section{Introduction}
Modern education policies around the globe have been rapidly adopting Problem Based Learning techniques in their education curriculum. The effectiveness of the PBL approaches in improving learning abilities has been extensively studied by several researchers and educators in the past. PBL is a constructive teaching model that promotes thinking, learning and solving problems at the same time\cite{thomas2000review}. Such skills are considered to be essential for the next generation of young learners to comprehend challenging global problems in the world. However, such skills are too frequently not acquired in traditional university syllabuses and are mostly popular in subject specific areas such as engineering and medical schools\cite{boud1997challenge}. There has been a significant push by governments and education policy makers all around the world to implement PBL methods to improve the skills of students at all levels of education stages (early childhood to graduate schools)\cite{trilling201221st,bell2010project}. Despite the progress and efforts there has been a significant delay in implementing such programs in university curriculum specially at graduate level. Moreover, the trend has been very slow in developed countries such as Japan where the implementation has mainly restricted to medical schools \cite{kozu2006medical}. While PBL techniques accentuates on improving problem solving aspect of student's learning, the notion of problem finding or problem discovery is neglected at most. Most problems that students deal with are perfectly set up for them, where the problems are well-defined with initial state, given clear goals and a finite number of operators to find the final solution \cite{chi1985problem}. Problem finding is an important component of creativity\cite{chand1993problem}. Though discussed in different studies in the field of psychology and education in the past\cite{getzels1975,csikszentmihalyi1970concern,gallagher1992effects,okuda1991creativity,zydney2008cognitive}, the definition of problem finding varies by point of view. Some considered it to be a cognitive strategy for effective learning \cite{graesser1992questioning}, others have defined it as cognition development or a process of cognition \cite{rosenshine1996teaching,chand1993problem}. Many have pointed the importance of engaging students in complex, unseen and ill-structured problem solving cases, that enable the students to perceive idealization of creative and meaningful relevance of their learnings and infering such knowledge in practical situations. Yet, there has been very few studies stressing the need to include ill-structured problem solving in classrooms. It has also been found that problem finding has strong influence on improving the creativity and overall problem solving process \cite{han2013influence}. 

The need for creative problem finding specially at graduate level is the main focus of this research article. Graduate school education aims at producing young graduates with definite skill sets and solving complex real-world problems. It corresponds to drastic changes in international society such as ``globalization" including development of human resources to nurture competencies such as transferable skills and to advance their specialty. However, if closely observed most graduate school curriculum does not focus on wicked problems or problems that are open ended with no good solution. Such wicked problems are mostly complex real-world problems that needs creative thinking and solving. On the other hand they are trained to focus on the solution aspect of the problem while completely ignoring the problem formulation process which is the most important step towards effective problem solving. Our study aims at discussing the need to include such ill-structured and wicked problems in graduate schools to produce better learners and problem solvers. We discuss our special graduate school program called the ``Nitobe School" (NS) program at Hokkaido University, Japan and, our course structure for training students to handle complex real-world wicked problems, and how it improves their overall creativity. We discuss NS framework of PBL training in context to divergent thinking and team based learning to handle global and local problems. 

%
%
%
%
\section{Nitobe School- Introduction}
Hokkaido University (founded 1876) is a leading research university in Japan of national importance and with over 18000 enrolled students including over 6000 graduate school students in 18 graduate schools (including professional graduate schools). To commemorate the 150th anniversary of the university, a future strategy was planned out to produce graduates that can posses sound judgment and deep insight and have ability to work in a inter-disciplinary, multicultural environment and play a leading role in the development of global society. In line with the above strategy, in 2015 a new trans-graduate school program named ``Nitobe School" came into existence under the ``Top Global University Project"  initiated by the MEXT (Ministry of Education, culture, sports and technology) Japan in Hokkaido University to propel the advancement and reformation of the university education system in Japan. Furthermore the aim is to create brilliant scholars that are trained with knowledge essential to handle real world problems and contribute to the advancement of the society. 
\begin{figure}[!b]
\centering
\includegraphics[scale=0.8]{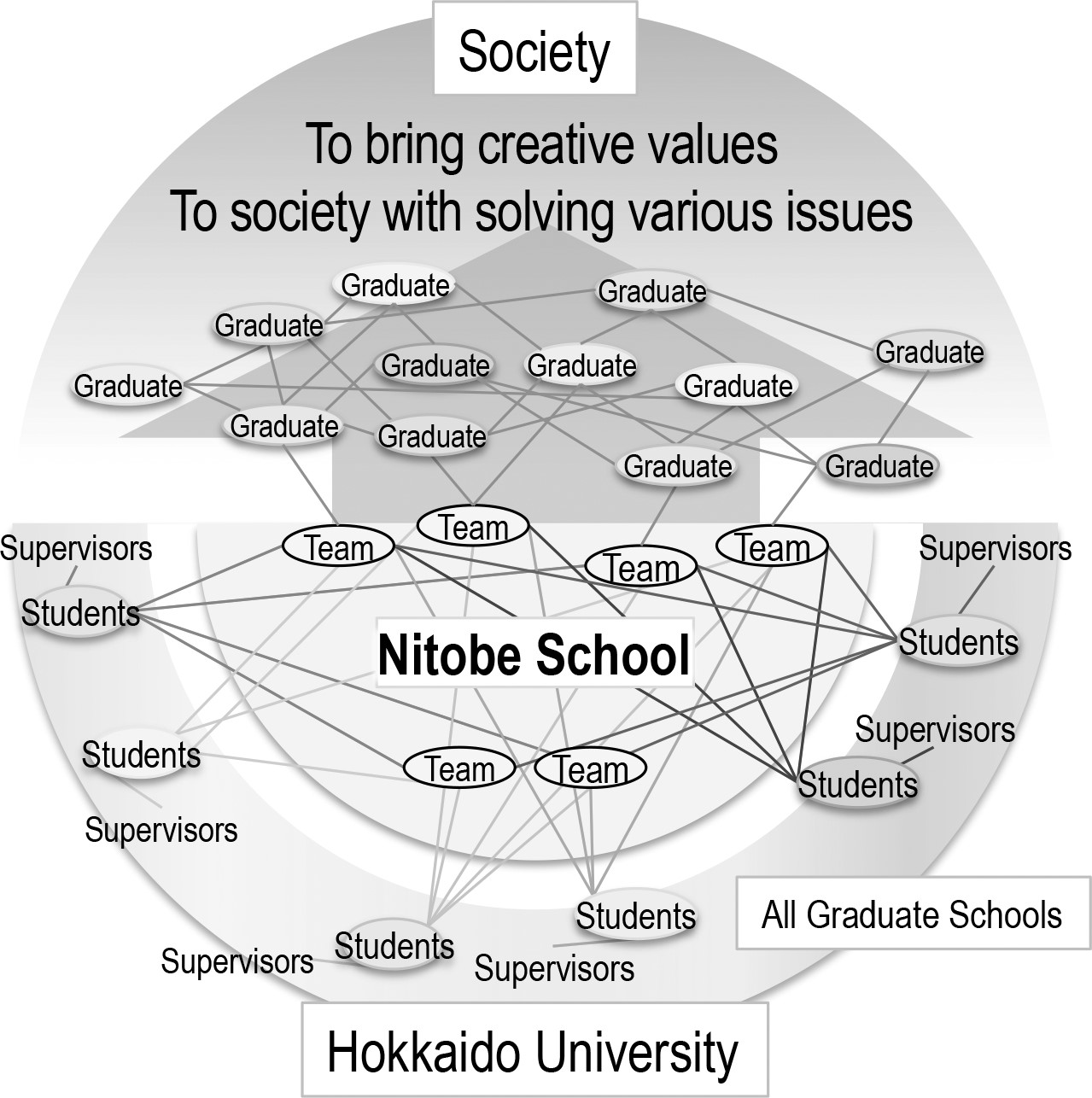}
\caption{Socio-Academic Ecosystem connecting Hokkaido University to the international society and disciplines. Figure from \cite{yamaDSIR, yamaIIAI} }
\label{fig:socio}
\end{figure}

\begin{figure}[!t]
\centering
\includegraphics[scale=0.8]{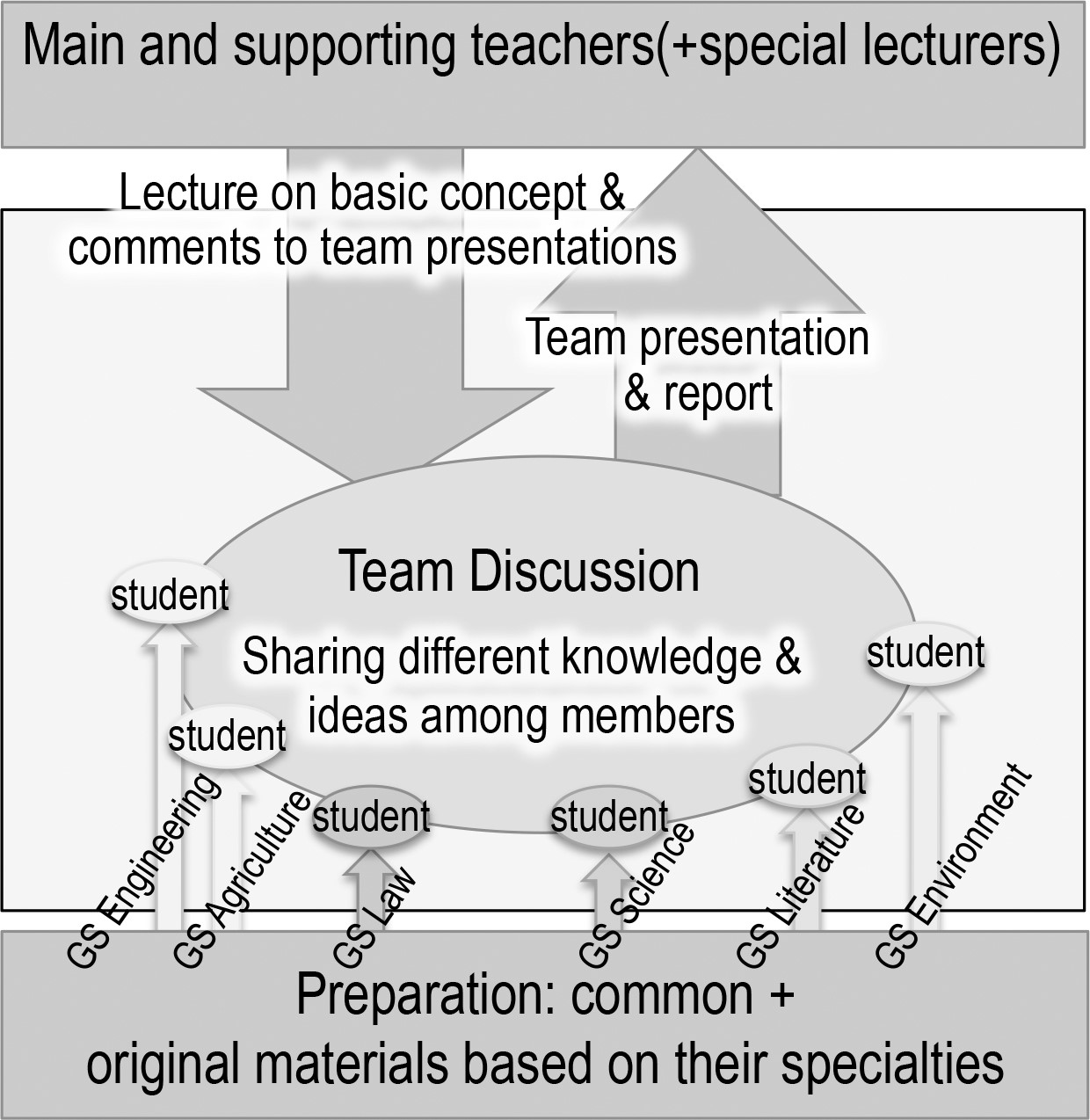}
\caption{Team based learning system at Nitobe School. Students from different graduate schools share their knowledge and learn new skills. Figure from \cite{yamaDSIR, yamaIIAI} }
\label{fig:TBL}
\end{figure}

 The aim of NS is to  develope high competencies in graduate students coming from each graduate school, and gives them the opportunity to learn beyond their field of expertise in an active learning and PBL style lecture. The overall goal is to provide world class education to young graduate students and provide skills that are aimed towards 21st century. Furthermore, Nitobe School will become a gateway of society, and students and graduates will produce “Socio-Academic Ecosystem” connecting Hokkaido University and international society over different disciplines (Fig. \ref{fig:socio}). The style of teaching at Nitobe School is primarily PBL and Team-Based Learning (TBL) system where students from different graduate schools are mixed to form a team of 5-7 students with multiple tutors in classroom (Fig. \ref{fig:TBL}) \cite{yamaDSIR,imai2016discussion, shimamura2016does}. A total of 18 graduate schools participate in the program. The NS has a basic program (one year duration) and an advanced program (one year duration). The basic program is divided into 4 quarters in an year with total intake of 120 students. Each course has three classes with around 30 students for each with several join-sessions. Each class has responsible instructor with an accompanying faculty and teaching assistants (TA's) for facilitation. The courses aims at improving students ability to learn in a multi-disciplinary environments along with international students to gain expertise in different skill-sets semester by semester.
\begin{table}[ht]
\centering
\renewcommand{\arraystretch}{1.5}
\begin{tabular}{l l l}
\hline
\textbf{Duration} & \textbf{Course Schedule} \\ 
\hline
March - April & Orientation for prospective students \\
\hline
May - June	& (Spring Term) “Basics of Team-based Learning” (Compulsory)\\
\hline
June - August	 &(Summer Term) “Practice of Team-based Learning”\\
\hline
Sep.-  Nov. & 	(Fall Term) “Problem Solving” \\
\hline
Nov. - Jan.	& (Winter Term) “Problem Finding” \\
\hline
\label{table:schedule}
\end{tabular}
\caption{Nitobe School Course Schedule}
\end{table} 

 The courses are (a) Basics of Team-Based Learning (Spring term), (b) Practice of Team-based Learning (Summer term), (c) Problem Solving (Autumn term) and (d) Problem Finding (Winter term) in order. Table 1,  shows the duration of  basic courses at NS in a calendar year. The course instructions are entirely given in English with flexibility to do discussions in Japanese and English. To support and improve student's ability to communicate and participate in discussions held in English, students of NS can take additional ``Nitobe School English" course throughout the year. The program is well supported by e-portfolios to monitor student's progress throughout the year\cite{imai2017utilization}. Upon each course completion, the student is credited 2 points and total 8 credits are required to complete the NS program. The instructors at NS also comes from diverse background and expertise, bringing their specialty and core knowledge into discussion. They are young faculties at Hokkaido University and the program provides a platform for faculty development for such young researchers to enhance their scope and to learn PBL style teaching. Further discussions in subsequent sections will be reserved to the Problem Finding course to keep the relevance to the topic. A detailed overview of the Nitobe School concept and origin can be found in work by Yamanaka et. al. \cite{yamaIIAI}.
 \subsection{Framework of Problem Finding course}
 The final course of the NS basic program is the Problem Finding (PF) course (winter term). As can be noted, the problem finding is preceded by Problem Solving course and it is believed that the student's have gained enough knowledge towards problem solving, project management and team based learning by the time they start the problem finding course. The reason for teaching PF as the last course is also due to the challenges the course offers where the students have to deal with real-world ill-structured problems. So when the students enroll for the  problem finding course, they already have sufficient knowledge to work in teams, handle small projects, gain enough leadership skills and manage group discussions. The course was offered 8 times weekly for 180 minute lecture per class. An important aspect of the PF course is that students should work in teams and work on real-world problems. To facilitate that, students are required to do field-work locally (around Hokkaido region) and select relevant topic related to the hidden problem they are finding. The field-work has to be done in team (5-7 student per team) and at least for 3 times out of the 8 week time frame. The significance of the field work will be discussed briefly later. A detailed course syllabus can be accessed from the Nitobe School website.

\section{Problem Finding Paradigm}

The PBL framework generally consists of the following characteristics (1) learning is student centered, (2) learning process occurs in small groups, (3) role of teachers are as guides and  facilitators, (4) problems form the organizing focus and stimulus for learning, (5) problems are a vehicle for the development of problem-solving skills and (6) new information is acquired through self-directed learning \cite{barrows1996problem}.  While (1)-(3) constitute the process, (4)-(6) deals with the aspect of problem that involves two components: representation and the solution process. The representation phase consists of the solver's interpretation or understanding of the problem. This is also called as defining the problem. It is essentially the first step towards effective problem solving.  The way problems are presented influences the student's problem-solving at large. If the problems are presented in a well-defined manner, students will solve the problem by going through a step by step method that can be identifying correct formula, putting data into the formula and then solving for the unknowns. However, for ill-structured or wicked problems have no specific formulations. There are no stopping rules; solutions to such problems are not true or false and there is no ultimate test of a solution to tackle wicked problems. Moreover, wicked problems can be considered to be born out of another problem and when working with wicked problems the planner has no right to be wrong\cite{beckman2012teaching,gallagher1992effects}.  Most ill-structured problems are mainly derived from real world events. Our aim therefore was to engage students with such real-world problems that are complex enough so that the students work together and rely on each other to solve them. Such challenging problems engages students effectively in relevant settings allowing them to bridge the theoretical knowledge with actual application\cite{edens2000preparing,ravankar2016nurturing}. This is not possible in traditional classrooms since students cannot comprehend the reality of existence of the problem. Therefore, framing and defining the problem plays an important role in the problem formulation process. PBL begins when student write the problem definition statement, build right hypothesis towards the problem, investigate the problem and finally solve the problem.  This involves students to use their domain knowledge skills and at the same time rely on team members to iteratively formulate the problem. During the process their hypothesis changes with each new observation that further validates the problem representation and formulation. A key element when dealing with ill-structured problem is student ownership of the problem. As students in group approach the problem naturally by various means such as questionnaire, interview, experiments, observation etc.  A significant scaffold to improve this process can be done by facilitators who can mentor and probe the right approach to attack the problem. An essential element to bring about the best out of the team would be to conceptualize the problem by brainstorming about linkages to the problem layers\cite{ankit2017problem}. Although the problem is mainly student oriented, the instructor can pave certain paths for the students to steer them into right directions that involves higher creativity and cognition by listing alternatives\cite{imai2016discussion,imai2017effect}.  Such listings can help students to understand things to do; things they already know and things they need to know. Group discussions and brainstorming plays a very important role to enhance this step. As student investigate, gather and share new knowledge, the  things to know column gets updated continuously. As more information is added, it replaces the things already known. This process also brings about intuitiveness in students and improve their creative thinking skills. A very similar model can be described in a multi-step model of recognition and redefinition as explained by Smith in \cite{smith1989defining, basadur1994new}. This divergent approach suggests that the problem's existence be challenged.  The several phases of divergent-convergent thinking involves identifying stakeholders, generating alternate perspectives, learning about the problem and creating the working definition of the problem. This is followed by explorations, where the problem is divided into sub-problems and explore possible causes of each sub-problems. Once the new information is added in the divergent-convergent model,  mistakes are identified and hypothesis are corrected.  Questions such as ``Why are we doing this" or ``How can we improve this" and ``What is the significance of this" nurtures originality and ideation. The process of ideation generates a large number of ideas to chose from. Students can record, make notes of their observation to assist in the process of idea creation. Such creative thinking makes students to try different perceptions and alternatives to the same problem, giving a 360 degree view of the problem. Such views are not derived from each other but are independently produced. Such problem discovery and exploration brings about the best creativity in students. Creativity as such is hard to teach, but in the PBL context when combined with problem finding and problem solving, creative thinking brings the best out of teams. 

\section{Evaluation}
We present the evaluation of the problem finding course for the academic year 2017-18.  During the course the students were asked to do field work locally and were given complete freedom to chose the topic of their choice. A theme was selected for the class in order for the students to give a start. The themes were different for each class and proposes an open ended problem such as (Improving Entrepreneurship in Hokkaido region of Japan, and Food Diversity and Gentrification). Such real-world problems were not faced by students before. During the course students were also trained the process of field-work. They were given instructions on how to contact stakeholders through emails and telephone, make appointments, prepare questionnaire for interviews, how to do the interview, how to record data when doing investigation and finally analyzing and exploring data to figure out meaning from the data. Their task was to formulate an ill-structured problem into a well-defined problem by going through the complete process. The facilitators responsibilities were to participate in the group discussion but not to influence students thinking and individual/group approach towards problem finding. However, they were instructed to provide the groups right directions if they found the discussions to be going  out of topic. The facilitators were also responsible in giving suggestions to improve the groups hypothesis every week and suggestions to make their investigation effective using tools such as questionnaire making, interview etiquette, note taking etc. Finally, the students were asked to present their findings in the final week in form of group presentations and by submitting a group report.

\subsection{Findings}
A questionnaire was prepared at the end of the course to evaluate the teaching and student understanding of the subject. From the total of 46 students, $N=26$ students responded the questionnaire. The questions ranged from evaluating students understanding of the problem finding as well overall course. With the scale, 1= \textit{strongly disagree} to 5= \textit{strongly agree}, almost all the students found the course to be relevant, worthwhile and useful for their future and career ($M=4.58$, $SD= 0.58$). Majority of the students agreed that they understood the notion of problem finding after the course ($M=4.69, SD=0.54$). With scale 5= Excellent to 1= Bad,  most student rated the course as excellent ($M=4.52, SD=0.58$). As compared to the previous years (2016-17) experience, the students rated the course higher ($M=4.30$ vs $M=4.52$). 

For the creativity, majority of the students believe that the course improved their creative thinking skills ($M=4.51, SD = 0.58$) and students felt that team discussion  improved their understanding of the problem and generated creative ideas ($M=4.65, SD=0.48$).

We also found several problems with the course that were reflected in the evaluation. For some students it was difficult to understand ill-structured problems due to its ambiguous and changing nature and that created difficulties in starting the field-work. Also, some students felt that they were frustrated due to the team balance and few members were not so active in proper team discussion and activities. Some of the students also felt that 8 weeks were not enough to understand problem finding process and perhaps the course should be 10 weeks or 12 weeks long so that longer field-work can be done. While half of the students had previous experience of doing field-work, most students stated that doing field-work to understand problem finding process was a good experience ($M=4.65, SD=0.48$). At the end of the course, student stated that ``they were uncomfortable with the ambiguity of the open ended problem and had no clue whether their approach was on right track or not, but by the end of the course they were satisfied with the problem resolution and team effort". Most student also agreed that they found the problem finding step difficult than the problem solving step ($M=4.11, SD=1.10$). 
\subsection{Discussion and final thoughts}

The problem finding course at NS gave us an important clue as to why it is important to teach problem formulation in PBL style of teaching. Such learning in real-world, authentic scenarios provides student to learn how to apply their creativity and knowledge to real problems. Such real problems have their relevance in different fields and require special skills and training. It provides students a direct link between theory and application giving them abilities to apply the inert knowledge to real problems when they start their professional career. There are many lessons learnt from the course. Firstly, for the instructors it was difficult to understand whether the students were progressing from the PBL during the first few weeks. Most students looked confused and it seemed that extra efforts and discussions were necessary to keep the students motivated. Secondly, for students who never had field-work experience, it was a steep learning curve in a short period of 8 weeks. But we also observed that when PBL takes place in teams, students learn quickly from their team members experience. It was also essential for the facilitators not to suppress students enthusiasm for challenging projects. But it was also viable to draw a line when required in order to keep the class on schedule and not to leave anyone behind. In this regard, including problem framing is exhaustive for the instructors as the uncertainty prevails and it is difficult to keep all teams at same level all the time. An important learning from the course was student's own feeling of responsible learning and working in teams. Working on real problems with members from diverse background and nationality brought a lot of creative ideas for discussion and kept everyone motivated which was very positive to see. From the field-work, we observed student's to hesitate in making the first contact with the stakeholder, but eventually the confidence was up when their investigation started. Fiel work also provided the students a small opportunity to manage the project within the time frame and we believe that knowledge gained from previous terms at NS helped student to carry out field-work smoothly. Finally, we felt that the course was very satisfying to teach.  With our current PBL model we believe that it will provide better opportunities for all to learn and enhance our skills. 
\section{Conclusion} 
We present the Nitobe School framework for graduate school education in a problem based learning scenario where problems are at the core of learning. We presented why it is important to include problem formulation and exposing students to ill-structured problems during the problem solving process. Such training invokes student's creative thinking bringing out their best when doing problem solving. The students learn to apply their inert knowledge and link theory to actual application. From our finding we believe that such inclusion in the PBL model where ill-structured problems act as stimulus for actively engaging students in inquiry and prepare better problem solvers. We found from our study that it enhances student's creative thinking. Such training is viable to meet the challenges of the next generation and produce graduates who are not only skillful but also ready for the twenty-first-century workplace.
%
%

\section*{References}

\bibliography{mybibfile}

\end{document}